\documentclass[runningheads]{llncs}

\usepackage[utf8]{inputenc}
\usepackage{graphicx}
\usepackage{hyperref}
\usepackage{amsmath}

\title{OR-UNet: an Optimized Robust Residual U-Net for Instrument Segmentation in Endoscopic Images}
\titlerunning{OR-UNet}

\author{
Fabian Isensee\inst{1,2} \and
Klaus H. Maier-Hein \inst{1}
}
\authorrunning{F. Isensee and Klaus H. Maier-Hein}
\institute{Division of Medical Image Computing, German Cancer Research Center (DKFZ), Heidelberg, Germany \and
Faculty of Biosciences, University of Heidelberg, Heidelberg, Germany}
\begin{document}

\maketitle  
\setcounter{footnote}{0}
\begin{abstract}
Segmentation of endoscopic images is an essential processing step for computer and robotics-assisted interventions. The Robust-MIS challenge provides the largest dataset of annotated endoscopic images to date, with 5983 manually annotated images. Here we describe OR-UNet, our optimized robust residual 2D U-Net for endoscopic image segmentation. As the name implies, the network makes use of residual connections in the encoder. It is trained with the sum of Dice and cross-entropy loss and deep supervision. During training, extensive data augmentation is used to increase the robustness. In an 8-fold cross-validation on the training images, our model achieved a mean (median) Dice score of 87.41 (94.35). We use the eight models from the cross-validation as an ensemble on the test set. 
\end{abstract}

\section{Introduction}
The development of computer and robotics-assisted interventions heavily relies on the detection and segmentation of surgical instruments in endoscopic images. While previous challenges already addressed this problem to some extend, the limited size of the provided datasets did not allow the development of sufficiently robust algorithms. The Robust-MIS 2019 challenge\footnote{\url{https://www.synapse.org/\#!Synapse:syn18779624/wiki/}} specifically adresses this shortcoming by providing the largest publicly available database to date with 5983 manually annotated images in the training set. The challenge is subdivided into three parts: 1) binary segmentation, 2) detection and 3) multiple instance segmentation of the surgical instruments.

Due to the required annotation effort, semantic segmentation datasets are typically much smaller than their image classification counterparts. While segmentation networks can use the data more effectively due to the dense nature of the predictions, smaller datasets still suffer from a substantially weaker representation of the natural variation in the images. For this reason, semantic segmentation in typical computer vision tasks is most often tackled by re-using ImageNet pretrained encoders in segmentation models \cite{chen2017deeplab}. Naturally, this paradigm has also been applied for the segmentation of surgical instruments in endoscopic images \cite{allan20192017}. 
However, recent evidence suggests that the benefits of using pre-trained encoders may be much smaller than initially thought, especially when used with larger target domain datasets \cite{he2018rethinking}. For this reason, all models presented in this paper were initialized randomly and trained from scratch using only the training data provided by the Robust-MIS challenge.

In this paper we describe our approach to the binary segmentation task. While we also submit predictions for tasks 2 and 3, it should be noted that these were generated \textit{ad hoc} using connected component analysis and our algorithm was not specifically designed nor tuned for these tasks.

\section{Method}
\subsection{Data}
Our model is trained solely on the training images provided by the Robust-MIS challenge. The challenge provides a total of 5983 manually annotated training images, stemming from 16 surgeries representing two different types of intervention (Prokto and Rectum, 8 surgeries each). Alongside each raw image, Robust-MIS also provides the 249 frames that precede it in the actual video of the surgical procedure. 

We do not make use of external data and do not use pretrained networks, either. While the aforementioned time information is certainly valuable for resolving particularly hard cases, our segmentation method only uses the query image as input to generate the segmentations.

\subsection{Preprocessing}
Since smaller image sizes allow for faster training and thus for more iterations in developing our model we process all images on half resolution. Given that the training images suffer from visible compression artifacts, we do not feel that a decrease in resolution will adversely affect our results. Thus, all images were resampled from an original size of 540 $\times$ 960 to a size of 270 $\times$ 480 pixels. Images were normalized by dividing with 255, resulting in intensity values in the range of $[0, 1]$.

\subsection{Network architecture}
\begin{figure}[t]
    \centering
    \includegraphics[width=1.0\textwidth]{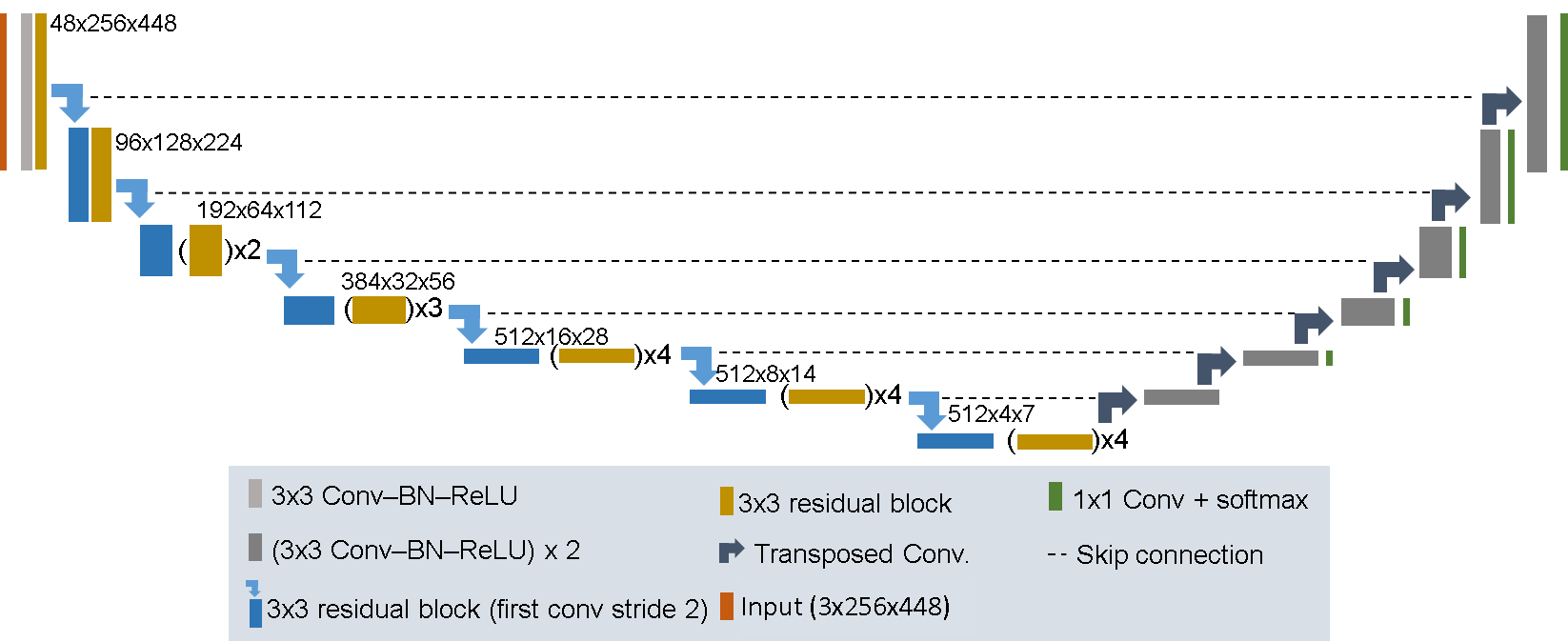}
    \caption{Network architecture. We use a 2D U-Net architecture that employs residual blocks in the encoder. Pooling operations are done via strided convolutions, upsampling is done with convolution transposed. We increase the number of residual blocks per resolution as the resolution decreases. The decoder constitutes of plain convolutions only. We generate segmentations at several resolutions throughout the decoder and use these segmentation maps to inject gradients deep into the network.}
    \label{fig:architectures}
\end{figure}

Our network, as depicted in Figure \ref{fig:architectures} is a 2D \cite{ronneberger2015u} U-Net that uses residual blocks in its encoder. The residual blocks follow their original formulation (see \cite{he2016identity}). The decoder has two convolutions per resolution, where each convolution is followed by batch normalization and ReLU nonlinearity. Aside from the segmentation layer at the end of the U-shape, we include additional segmentation layers at lower resolutions that are intended to improve the gradient flow through the network. At the highest resolution, the network has 48 feature maps, a number that is doubled whenever the resolution is decreased, up to a total of 512 feature maps. We do not go beyond 512 feature maps for practical reasons: even with this limit in place, a single model checkpoint file is as large as 1GB.

\subsection{Network Training}

\subsubsection{Loss Function}
We train our model with deep supervision. At the highest output resolution, the loss is simply the sum of the soft Dice loss \cite{drozdzal2016importance} and the regular pixel-wise cross-entropy loss function.

For predictions at lower resolutions we create downsampled segmentation maps which were generated by average pooling a one hot encoding of the ground truth. Average pooling generates soft segmentations (not 0-1 encodings). For these predictions, we use the sum of the mean squared error loss function with a variant of the Dice loss that was adapted for soft ground truth labels. 

This variant of the dice loss is computed as follows: Let x and y be the predicted softmax probabilities and the soft ground truth, respectively. Both have shape $(b, c, X, Y)$, where $b$ is the batch size, $c$ the number of classes (here 1) and $X$ and $Y$ are the spatial dimensions. 
We can then define true positives (TP), false positives (FP) and false negatives (FP) as:
\begin{equation}
    TP = max(0, y - |x - y|)
\end{equation}
\begin{equation}
    FP = max(0, x - y)
\end{equation}
\begin{equation}
    FN = max(0, y - x)
\end{equation}

Note that $max()$ is the elementwise max operation. We then sum TP, FP and FN over the spatial dimension as well as the batch dimension to obtain an aggregated scalar value for each. These scalar values, denoted $tp, fp$ and $fn$ can then be used to compute a dice loss:
\begin{equation}
    loss_{dice} = - \frac{2 tp}{2 tp + fp + fn}
\end{equation}

The total loss of the network is the weighted sum of the losses at different resolutions. Losses at lower resolution have lower coefficients ($\frac{1}{2}$ for half resolution, etc) and all coefficients are normalized so that they sum to 1. 

\subsubsection{Data Augmentation}
Despite the rather large size of the dataset we observed substantial overfitting in our initial experiments. To improve the generalization of our network we make use of of the following augmentations: rotations, elastic deformations, scaling, mirroring, additive Gaussian noise, brightness, contrast and gamma.

All augmentations were implemented with the \textit{batchgenerators} package \footnote{\url{https://github.com/MIC-DKFZ/batchgenerators}} and are applied on the fly during training

\subsubsection{Training Procedure}
The model is trained in an 8-fold cross-validation where, for each surgery type, we leave one surgery out per fold. For each fold, this results in 14 surgeries for training and 2 for evaluation. Training is done on randomly selected patches of size 256 $\times$ 448 pixels. We train with a batch size of 64 for 2000 epochs, where we define an epoch as the iteration over 100 batches. 
We learn the parameters of the network via stochastic gradient with nesterov momentum (momentum=0.9) and an initial learning of 1. During training the learning rate slowly decays to 0 with the \textit{polyLR} schedule \cite{chen2017deeplab}.

We use make use of mixed precision training as provided by the Nvidia \textit{amp} package\footnote{\url{https://github.com/NVIDIA/apex/tree/master/apex/amp}} to speed up the training and reduce the amount of GPU memory required to train the model. The training of a single model takes about 4 days on a Nvidia V100 GPU (32GB).

\subsection{Inference}
For inference we make use of an ensemble that consists of the eight models obtained from the training set cross-validation. The disparity between patch size (256 $\times$ 448) and image size (270 $\times$ 480) is resolved by predicting the images with a sliding window approach. Softmax predictions from the ensemble members are averaged for each image and subsequently upsampled to the original image size (540 $\times$ 960). The binary segmentation is then obtained via argmax.

The detection task for this challenge is formulated such that it also accepts the output of an instance segmentation algorithm. Since our model is unable to create instance segmentations, our submission to the detection and multiple instance segmentation tasks is created \textit{ad hoc} by using connected component analysis on the segmentation maps. Hereby, each isolated foreground object is assigned a unique ID.

\section{Results}
\begin{figure}[t!]
    \centering
    \includegraphics[width=0.9\textwidth]{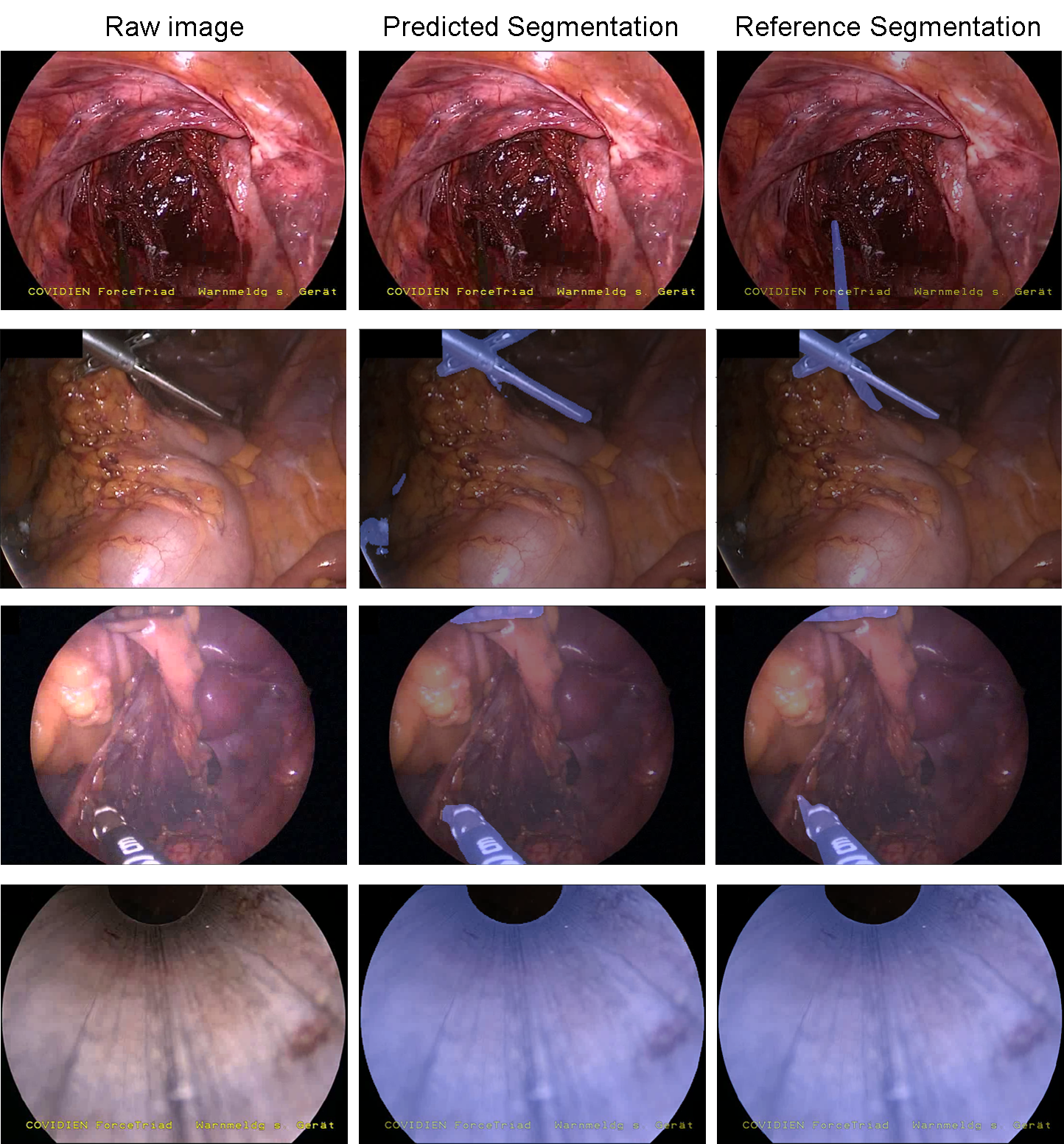}
    \caption{Qualitative segmentation results. The columns show the raw image, the predicted segmentation and the reference segmentation from left to right. Segmentations are shown as overlay to facilitate interpretation. The top row shows one of the worst results (Dice = 0), the first show shows the 10th percentile (Dice = 76.00), the second row the median (Dice = 94.35) and the bottom row shows the best result (Dice=99.82). All results are taken from the training set cross-validation.}
    \label{fig:results}
\end{figure}

Our model achieved an average Dice score of 87.41 (median 94.27, IQR = 89.95 - 96.4) in the eight-fold cross-validation on the training data. This score was obtained by computing the dice score for each image, followed by averaging over all training images. Note that our score computation also included images with no instrument present. For these cases, the Dice score is 0 if a single false positive pixel is predicted by the network. If both the prediction and the ground truth are empty, the image is excluded from averaging (the Dice is undefined in this case). Thus, the Dice score we report here is lower than it would be if the evaluation were done by the challenge organizers. Note that for the evaluation of the test set, images will be excluded if they do not contain a ground truth label.

Figure \ref{fig:results} gives a qualitative overview of the segmentation results. Its rows show the worst, 10th percentile, median and best result. 

\begin{figure}[t!]
    \centering
    \includegraphics[width=0.6\textwidth]{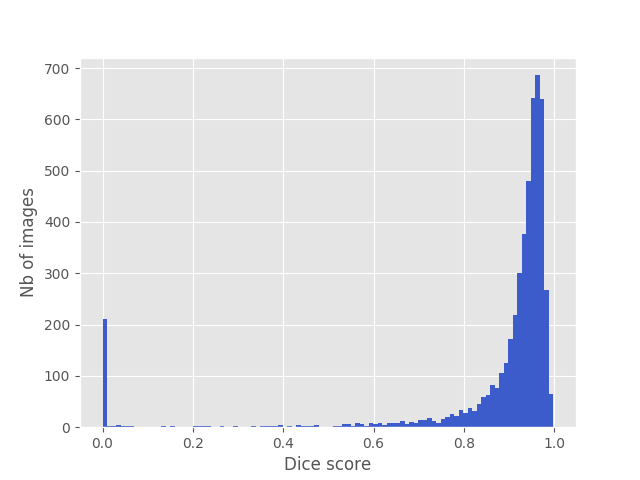}
    \caption{Histogram of the Dice scores on the training set cross-validation. The large number of images with a Dice = 0 is caused by images where either the network predicted a false positive label or where the network failed to segment a hidden instrument. The histogram underlines that the segmentation accuracy is very high for the vast majority of images.}
    \label{fig:hist}
\end{figure}

The histogram in Figure \ref{fig:hist} shows the distribution of Dice scores on the training cases. The large majority of the images is segmented accurately. The histogram illustrates the large number of Dice 0 images that drag down our mean Dice score. Most of these constitute cases where either the network produced false positive or false negative results only.

We did not compute metrics for the detection and instance segmentation task on the training cases.

\section{Discussion}
The results presented in the previous section demonstrate that the proposed model has a strong segmentation performance on the Robust-MIS dataset. While the median dice scores is quite high (94.35), the average dice score of 87.41 is substantially lower, mostly due to the large number of images where the model obtained a Dice score of 0 (e.g. by producing false-positive pixels when no instrument is present, or predicting no positive pixels if an instrument was present). As stated previously, the evaluation of the test set for the binary segmentation task will not include images with no ground truth present. This, taken together with the fact that we predict the test set with an ensemble of eight models, might result in a higher Dice score for our method on the test set.

We explicitly chose to not use a pretrained encoder for our model we followed the hypothesis that the dataset is large enough for self-sufficient training. An experimental backup of this decision is missing though. We expect other participants to choose pretrained networks and are excited to see how their results compare to ours.

During method development we experimented with including time series information in a very straightforward way: by stacking several frames from the time series in the color dimension (resulting in an input of more than 3 color channels). These experiments did not improve our results, which is why the present model only makes use of the query image for its segmentation. Other methods of including information from the preceding frames may prove to be more fruitful (for example in the form of recurrent neural networks) but were not explored in the context of this project.

While we did not explicitly design a model that can create instance segmentations and thus be used to participate in the detection or instance segmentation task, visual inspection of the training images revealed that the large majority of images includes only a single instrument. Even if multiple instruments are present, they do not touch in most cases. This led us to the idea of simply submitting our binary segmentation results to the other tasks after running them through a connected component analysis. This approach may deliver a solid baseline of what can be achieved on basis of just a strong segmentation performance.

 \section*{Acknowledgements}
 The authors would like to thank Gregor K\"ohler for proofreading the manuscript.

\bibliographystyle{splncs04}
\bibliography{main}
 
\end{document}